\documentclass[manuscript,nonacm, 10pt, oneside]{acmart}

\AtBeginDocument{%
  }

\makeatletter
\@twosidefalse
\@mparswitchfalse
\makeatother

\usepackage{pgfplots}
\DeclareUnicodeCharacter{2212}{−}
\usepgfplotslibrary{groupplots,dateplot}
\usetikzlibrary{patterns,shapes.arrows}
\pgfplotsset{compat=newest}

\usepackage{amsmath,amssymb,amsfonts,mathtools}

\usepackage{graphicx,xcolor,svg}

\usepackage[linesnumbered,ruled,vlined]{algorithm2e}
\usepackage{algorithmicx}
\usepackage{algpseudocode}

\usepackage{booktabs,multirow,tabularx,array}

\usepackage[T1]{fontenc}
\usepackage{times}
\usepackage{fancyhdr}
\usepackage{framed}
\usepackage{url}
\usepackage{balance}
\usepackage{caption}
\usepackage{subfigure}
\usepackage{upquote}
\usepackage{comment}
\usepackage{ragged2e}
\usepackage{academicons}
\usepackage{adjustbox}
\usepackage{hyperref}

\newcommand{\papername}{ACCESS-AV}

\newcommand{\arnab}[1]{\textcolor{blue}{\textbf{#1}}}

\newcommand{\rred}[1]{\textcolor{red}{\textbf{#1}}}

\renewcommand{\thefootnote}{\textit{\alph{footnote}}}

\newcommand{\supc}[1]{%
    (\tikz[baseline=(char.base)]{
        \node[shape=circle, fill=black, text=white, inner sep=0.5pt] (char) {\small #1};
    })
}

\DeclareRobustCommand{\circlednum}[1]{%
    \tikz[baseline=(char.base)]{
        \node[shape=circle, fill=black, text=white, inner sep=1pt] (char) {\small #1};
    }%
}

\begin{document}

\title{\texorpdfstring{\papername{}: {\underline{A}daptive \underline{C}ommunication-\underline{C}omputation Cod\underline{e}sign for \underline{S}u\underline{s}tainable \underline{A}utonomous \underline{V}ehicle Localization in Smart Factories}}%
{\papername{}: Adaptive Communication-Computation Codesign for Sustainable Autonomous Vehicle Localization in Smart Factories}}


\author{Rajat Bhattacharjya}
\affiliation{%
  \institution{University of California, Irvine}
  \city{Irvine}
  \state{California}
  \country{USA}}
\email{rajatb1@uci.edu}

\author{Arnab Sarkar}
\affiliation{%
  \institution{Indian Institute of Technology, Kharagpur}
  \city{Kharagpur}
  \state{West Bengal}
  \country{India}}
\email{arnabsarkar@kgpian.iitkgp.ac.in}

\author{Ish Kool}
\affiliation{%
  \institution{Independent Researcher}
  \city{Lucknow}
  \state{Uttar Pradesh}
  \country{India}}
\email{kool382@gmail.com}

\author{Sabur Baidya}
\affiliation{%
  \institution{University of Louisville}
  \city{Louisville}
  \state{Kentucky}
  \country{USA}}
\email{sabur.baidya@louisville.edu}

\author{Nikil Dutt}
\affiliation{%
  \institution{University of California, Irvine}
  \city{Irvine}
  \state{California}
  \country{USA}}
\email{dutt@uci.edu}

\renewcommand{\shortauthors}{Bhattacharjya et al.}

\begin{abstract}

Autonomous Delivery Vehicles (ADVs) are 
increasingly used
for transporting goods in  5G network-enabled smart factories,
with the compute-intensive localization module presenting a significant opportunity for optimization. 
We propose \emph{\papername{}}, an energy-efficient Vehicle-to-Infrastructure (V2I) localization framework that leverages existing
5G infrastructure in smart factory environments. 
By 
opportunistically accessing the periodically broadcast 5G Synchronization Signal Blocks (SSBs) for localization, 
\emph{\papername{}}
obviates the need for dedicated Roadside Units (RSUs) or additional onboard sensors to achieve energy efficiency as well as cost reduction. 
We implement an Angle-of-Arrival (AoA)-based estimation method using the Multiple Signal Classification (MUSIC) algorithm, optimized for resource-constrained ADV platforms through an adaptive communication-computation strategy that dynamically balances energy consumption with localization accuracy based on environmental conditions such as Signal-to-Noise Ratio (SNR) and vehicle velocity. Experimental results demonstrate that \emph{\papername{}} achieves an average energy reduction of 43.09\% compared to non-adaptive systems employing AoA algorithms such as vanilla MUSIC, ESPRIT, and Root-MUSIC. It maintains sub-30~cm localization accuracy while also delivering substantial reductions in infrastructure and operational costs, establishing its viability for sustainable smart factory environments.

\end{abstract}



\maketitle

\begingroup
\renewcommand{\thefootnote}{}
\footnotetext{\textit{\small Authors' version posted for personal use and not for redistribution.}}
\endgroup

\section{Introduction}
\label{sec:intro}
\begin{figure}
\includegraphics[width=0.9\textwidth]{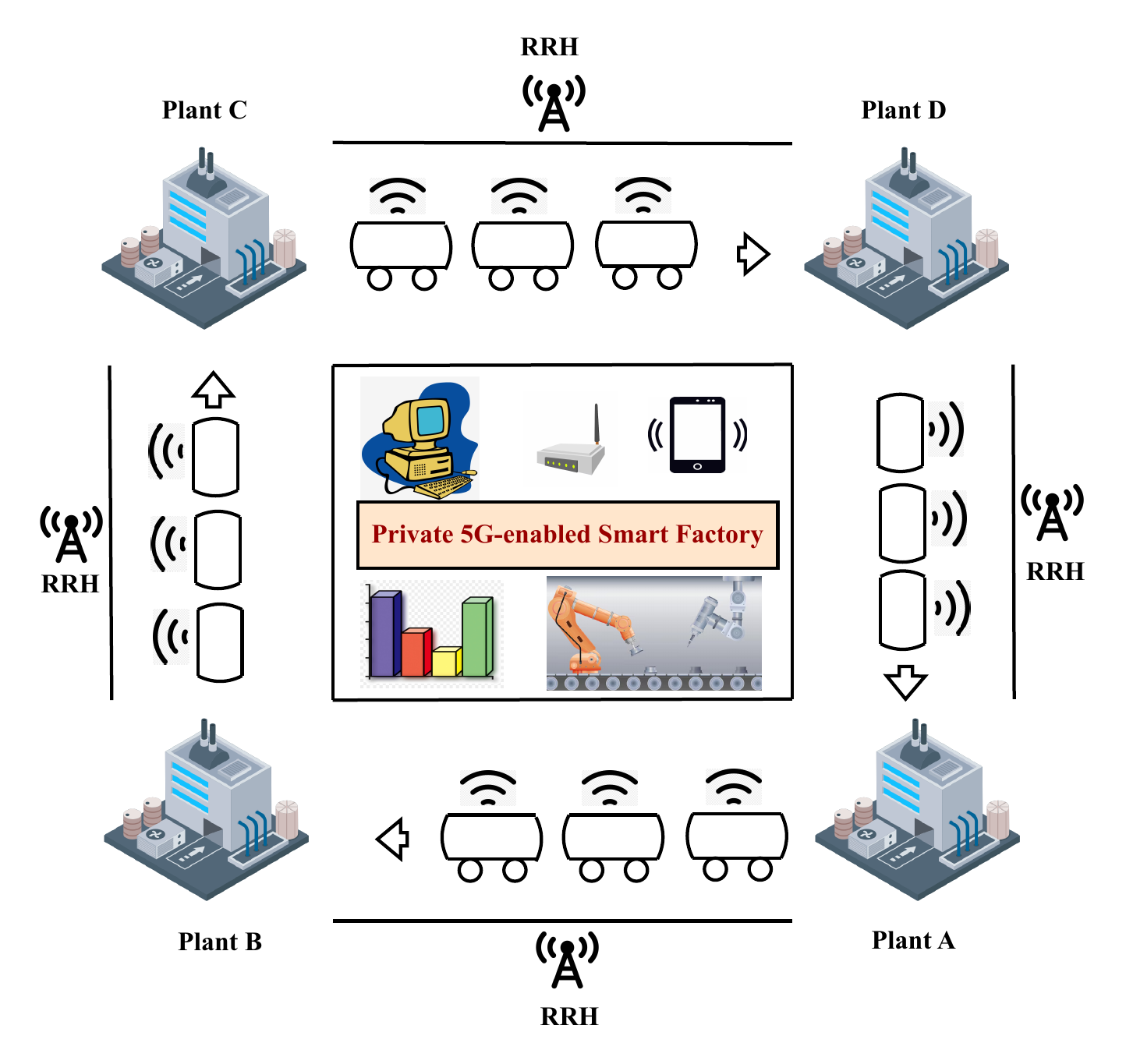}
\caption{A Smart Factory Infrastructure Powered by Private 5G: Autonomous Delivery Vehicles (ADVs) transport goods across plants while leveraging 5G Synchronization Signal Blocks (SSBs) from roadside Remote Radio Heads (RRHs) for localization.}
\label{fig:outline}
\end{figure}

With rapid advances in computing systems and robust network infrastructure, factories are increasingly transitioning toward autonomy. Modern smart factories, often equipped with private 5G infrastructure~\cite{pvt5g}, span vast areas covering several acres of land~\cite{hplc}. Safety regulations dictate significant separation between plants, with distances ranging from a few hundred meters to a half kilometer~\cite{separation}. These extensive layouts pose logistical challenges for the efficient and safe transportation of goods across facilities.

To address these challenges, \textbf{autonomous delivery vehicles (ADVs)} equipped with Level 4 (L4) autonomy have emerged as a promising solution to transport goods between plants~\cite{uisee}, as shown in Figure~\ref{fig:outline}. These vehicles reduce human dependency and improve operational efficiency and safety~\cite{factory}. However, the deployment of ADVs requires overcoming critical challenges related to \textit{both} onboard energy consumption (extending battery life for long-duration operations~\cite{adv}) \textit{and} overall system sustainability (reducing deployment costs and the environmental impact of additional infrastructure). Toward these ends, we explore energy-efficient localization strategies that leverage existing 5G networking infrastructure. Our goal is to simultaneously extend ADV operational durations and improve smart factory sustainability by eliminating the need for dedicated transmission infrastructure— also known as Roadside Units (RSUs)~\cite{rsu}— thereby significantly reducing both embodied and operational carbon footprints~\cite{jones,cc}.

We focus on localization, an integral perception module in an autonomous vehicle's computational stack~\cite{chauffeur,avstack}, which presents an opportunity to address the intertwined energy and sustainability bottlenecks in ADVs. Specifically, we focus on \textbf{Vehicle-to-Infrastructure (V2I) localization} using Angle of Arrival (AoA)-based estimation via the Multiple Signal Classification (MUSIC) algorithm~\cite{music_lite}. MUSIC provides high-resolution AoA estimates and has proven effective in diverse applications, from biomedical sensors~\cite{use2} and seismic monitoring~\cite{seismic}, to joint radar-communication systems~\cite{musicuse} and challenging multipath-rich V2I environments~\cite{italy}. Studies show that AoA-based localization can offer greater accuracy compared to GPS-based positioning~\cite{italy, italy2} and is well-suited for robust ADV localization~\cite{music_lite}.


Our approach offers a holistic and sustainable solution by primarily focusing on (1) \textbf{eliminating the need for dedicated Roadside Units (RSUs)}~\cite{rsu, aviral}. We achieve this by repurposing the factory's \textbf{existing private 5G infrastructure} for V2I communication, a foundational shift that drastically reduces deployment costs, complexity, and carbon footprint~\cite{jones}. Within this sustainable architecture, we employ an \textbf{adaptive communication-computation codesign} approach to dynamically balance localization accuracy and onboard energy consumption. This adaptivity is crucial, leveraging real-time factors like environmental quality (e.g., SNR) and vehicle velocity to ensure localization is both trustworthy and computationally efficient. Furthermore, our strategy enhances sustainability on the vehicle itself by (2) minimizing onboard hardware complexity and cost through reliance solely on a single wireless receiver, obviating the need for additional sensors (like cameras or LiDAR) for this task.

Figure~\ref{fig:outline} is a cartoon of a smart factory deploying a private 5G network that uses Remote Radio Heads (RRHs)~\cite{rrh} strategically positioned along roads interconnecting adjacent factory plants. These RRHs, part of the existing communication infrastructure, effectively serve as \textit{proxy RSUs} in our system. Our framework, \emph{\papername{}}, opportunistically exploits the 5G Synchronization Signal Blocks (SSBs)~\cite{ssb, alexandra} periodically broadcast by these RRHs to enable efficient V2I localization. \emph{\papername{}} leverages its adaptive communication-computation approach to dynamically calculate the AoA on the ADV's onboard system, achieving precise localization while minimizing energy consumption. Operating within the factory's predictable, closed-loop environment, \emph{\papername{}} ensures energy-efficient navigation along predefined ADV paths.

By specifically addressing the computational demands and energy-efficiency challenges of the MUSIC algorithm~\cite{music_lite} through this adaptive codesign, \emph{\papername{}} dynamically balances localization accuracy against onboard energy expenditure in real-time. By intelligently leveraging existing network resources (5G RRHs) and non-dedicated signaling (SSBs), the framework ensures accurate, scalable, cost-effective, and energy-efficient localization, supporting the broader vision of sustainable smart factories.

To the best of our knowledge, \emph{\papername{}} is the first work to jointly address ADV onboard energy constraints and infrastructure sustainability considerations through a cohesive, adaptive localization approach leveraging opportunistic V2I communication. \emph{\papername{}}'s main contributions are that we:

\begin{itemize}
\item Eliminate the need for dedicated RSUs and additional signaling by opportunistically leveraging periodically broadcast SSBs from existing 5G-enabled RRHs in the factory infrastructure. This approach reduces deployment costs, system complexity, and overall carbon footprint~\cite{jones}.
\item Simplify onboard hardware requirements by utilizing 
 only an Orthogonal Frequency-Division Multiplexing (OFDM) wireless receiver for localization, eliminating the need for additional sensors and their associated maintenance costs while promoting sustainable development.
\item Present a communication-computation codesign framework 
that dynamically balances localization accuracy and onboard energy consumption via an algorithm that adapts to real-time conditions.

\item Demonstrate 
average energy reduction of 43.09\% compared to non-adaptive systems employing vanilla MUSIC, ESPRIT~\cite{esprit}, and Root-MUSIC~\cite{root}, while maintaining high localization accuracy with worst-case error below 0.3 meters.

\item Further quantify the sustainability benefits of our approach via empirical cost estimation and analysis, which demonstrates over a $130\times$ reduction in onboard sensor costs.

\end{itemize}

The rest of this paper is structured as follows: Section~\ref{sec:background} provides relevant background; Section~\ref{sec:sys} details the system model; Section~\ref{sec:loc} presents the adaptive localization framework; Sections~\ref{sec:codesign}, \ref{sec:expt}, and \ref{sec:eval} discuss the communication-computation codesign framework, experimental setup, and evaluation, respectively; Section~\ref{sec:related} reviews related work; and Section~\ref{sec:conclusion} concludes the paper with insights and future directions. Acronyms used throughout this paper are summarized in the appendix for enhanced readability.

\section{Background}
\label{sec:background}
We choose an AoA-based localization strategy that leverages MUSIC for localization by opportunistically utilizing existing infrastructure. This choice is motivated by several factors. First, MUSIC is particularly well-suited for this application because its high-resolution AoA estimation capability enables it to reliably resolve multiple signal paths even in multipath-rich environments, ensuring fairly accurate localization in complex V2I scenarios~\cite{italy,music2}.
Second, AoA-based estimation methods have been shown to be more reliable than GPS-based positioning~\cite{italy,italy2}. 
Third,
this allows us to leverage the factory's existing 5G infrastructure for localization,
obviating the need for dedicated RSUs and thereby supporting sustainability.
Fourth, this reduces system costs and complexity onboard, as only a single wireless receiver is required for performing localization. 

However, it’s important to note that MUSIC is resource-intensive, mainly due to the computational demands of matrix decomposition techniques like eigenvalue decomposition (EVD) and/or singular value decomposition (SVD), which are power-hungry and computationally expensive~\cite{music_lite}. This poses a challenge for deploying MUSIC on resource-constrained embedded platforms, such as an onboard computing system in an ADV. To address this challenge, we propose \emph{\papername{}}, which adapts 
communication and computation in synergy to meet the goal of onboard energy-efficiency.

\begin{figure}
\includegraphics[width=0.6\textwidth]{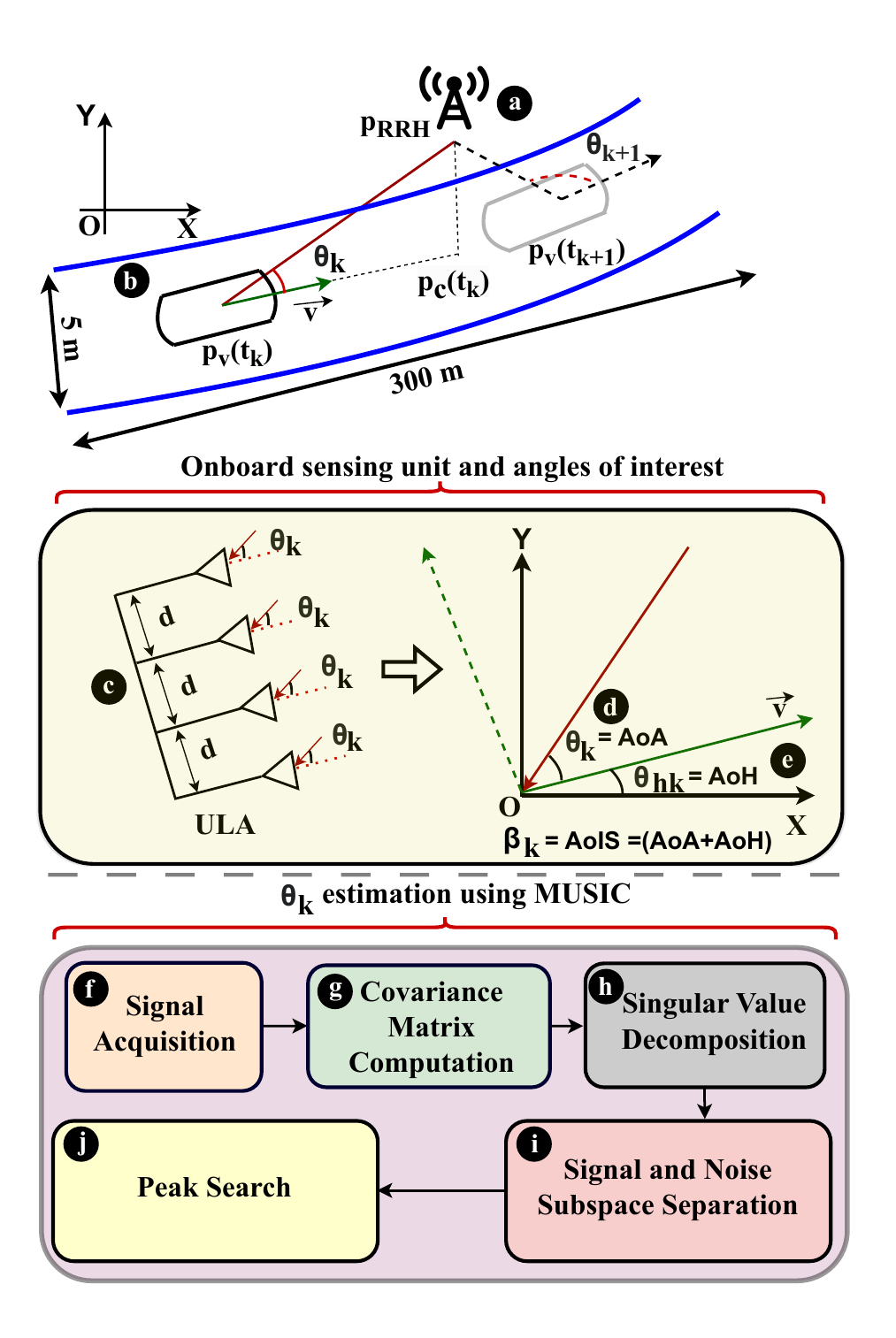}
\caption{Vehicle trajectory and track model with MUSIC flow used for AoA ($\theta_k$) estimation.~\circlednum{a} and ~\circlednum{b} illustrate the infrastructure and track model. ~\circlednum{c} depicts the onboard sensing unit, a ULA with four antennas. ~\circlednum{d} and ~\circlednum{e} present the estimated AoA and AoH, respectively. ~\circlednum{f} to ~\circlednum{j} outline the MUSIC processing steps for AoA estimation.}
\label{fig:specific}
\end{figure}

We now present a brief background on SSB and MUSIC:
\\
(1) \textbf{Synchronization Signal Block (SSB):} As illustrated in Figure~\ref{fig:outline}, ADVs travel along fixed paths within the factory, with each path covered by the cell of a single RRH. The ADVs localize themselves using the SSBs~\cite{sab} transmitted by the RRH serving their current cell. In 5G New Radio (5G NR), SSBs are generated by the gNodeB (gNB)~\cite{gnb} to facilitate initial cell detection, synchronization, and network access for User Equipment (UE). In deployments that utilize RRHs~\cite{rrh}, the gNB centrally manages SSB generation while the RRHs transmit these signals over the air to extend coverage and enhance connectivity. Each SSB comprises a predefined sequence of bursts and follows a network-dictated periodicity. Its structure includes the Primary Synchronization Signal (PSS) and Secondary Synchronization Signal (SSS), which together allow the derivation of the Physical Cell Identity (PCI) for unique cell identification, and the Physical Broadcast Channel (PBCH) that carries essential system parameters such as the Master Information Block (MIB) for UE configuration and access~\cite{ssb}.\\
(2) \textbf{Multiple Signal Classification (MUSIC):}
MUSIC is a high-resolution subspace-based algorithm used for estimating the Angle of Arrival (AoA) of signals impinging on an antenna array~\cite{music, music_lite}. As shown in Figure~\ref{fig:specific}, for a Uniform Linear Array (ULA) of antennas~\supc{c}, the MUSIC algorithm operates by first constructing a covariance matrix~\supc{g} from the received signals~\supc{f}. It then performs Singular Value Decomposition (SVD)~\supc{h} on this covariance matrix to separate the signal and noise subspaces~\supc{i}. Finally, MUSIC computes a spectrum by evaluating candidate steering vectors~\supc{j} against the noise subspace, identifying the direction of arrival by maximizing the MUSIC spectrum, mathematically defined as:
\begin{equation} \label{MUSIC_eq}  
P_{MU}(\alpha) = \frac{1}{a^{H}(\alpha) E_{N} E_{N}^{H} a(\alpha)}  
\end{equation}
Here, \( P_{MU}(\alpha) \) denotes the MUSIC pseudospectrum, \( E_{N} \) is the matrix containing eigenvectors spanning the noise subspace, and \( a(\alpha) \) represents the array steering vector corresponding to a signal arriving from direction \( \alpha \). The superscript \( H \) indicates the Hermitian transpose~\cite{herm}. 

Equation~\ref{MUSIC_eq} produces a prominent peak at the angle corresponding to the actual direction at which the signal impinges on the antenna array, denoted as \( \theta_k \) in Figure~\ref{fig:specific}. The estimated AoA is thus obtained by identifying this peak as:
\begin{equation}
\theta_k = \arg\max_{\alpha} P_{MU}(\alpha)
\end{equation}

\section{System Model}

\label{sec:sys}

\subsection{Track, Infrastructure, and Vehicle Trajectory Model}
While Figure~\ref{fig:outline} provides a birdseye view of the smart factory infrastructure, Figure~\ref{fig:specific} zooms in on a specific track segment connecting two plants. In this segment, an ADV navigates from the source plant to the destination plant, with a single RRH covering the corresponding cell by transmitting SSBs over the air.

To illustrate our~\emph{ACCESS-AV} approach, we consider typical track parameters representative of the route between two plants in a factory\footnote{The track measurements correspond to the separation distance between plants in a factory~\cite{separation} and are also inspired by various autonomous driving research test facilities~\cite{mcity, tecnalia}.}: 
the track is modeled as a horizontal single-lane path with a length of $300\,\mathrm{m}$ and a width of $5\,\mathrm{m}$, featuring a mild leftward bend with a cumulative angle of $14.32^\circ$.

From Figure~\ref{fig:specific}, we observe the RRH  positioned at $p_{RRH}$,
as determined from a pre-computed map, and always accessible to the vehicle. In a 2D Cartesian coordinate system with the vehicle starting from the origin, the coordinates of $p_{RRH}$ are given by $(205\,\mathrm{m},\,100\,\mathrm{m})$. Also, $p_v(t_k) = [x_v(t_k)\quad y_v(t_k)]^T$ (with $(\cdot)^T$ denoting the transpose operator) represents the position of the ADV at time instant $t_k$, and $\theta_k$ denotes the corresponding AoA. Without loss of generality, our analysis is restricted to planar localization.

Along the $300\,\mathrm{m}$ track, the ADV operates in three phases: acceleration, constant speed, and deceleration. It starts from an initial velocity of $0\,$kmph from the source plant, accelerates at a rate of $0.5\,$m/s$^2$ until reaching $15\,$kmph, and finally decelerates back to $0\,$kmph upon reaching the destination plant. These values are selected in accordance with safety standards for ADVs\footnote{These parameters align with the velocity range of state-of-the-art autonomous delivery vehicles~\cite{gen3}.}.

At time $t_k$, we define two angles of interest below:

\subsubsection{Angle of Arrival (AoA ($\theta_k$))} This is the incident angle from the RRH to the Uniform Linear Array (ULA) of antennas in the vehicle as shown in Figure~\ref{fig:specific}. The vehicle's onboard computing system estimates the AoA with the help of MUSIC at both fixed and dynamic time intervals which will be discussed in the following subsections.

\subsubsection{Angle of Heading (AoH ($\theta_{hk}$))} 
This represents the direction in which the ADV is heading with respect to the X-Y plane. 
AoH changes continuously as the vehicle moves along a curved path (e.g., 
 Figure~\ref{fig:specific}'s track).
To better understand AoH, consider the following scenario:
At time $t=0$, it's reasonable to assume the vehicle's position and direction are known. Using data from the onboard odometer, we can determine the vehicle's velocity ($v$) at this point. Without loss of generality, by applying a simple kinematic model, we can then calculate the vehicle's position at time $t=1$. However, if the vehicle changes direction at time $t=1$, the velocity must be decomposed into its corresponding $x$ and $y$ components to determine the shift in $p_{vx}$ and $p_{vy}$ (
the decomposed components of the position vector $p_v$ in X and Y direction respectively). 
Note that
a simple kinematic model will not work, as the velocities along $x$ and $y$ are not known separately. 
Therefore, it's critical to compute
the Angle of Heading (AoH) 
to 
accurately track the vehicle's new direction relative to the reference plane, and decompose the velocity components in each direction. 
Using this, we can calculate the next position for time $t = 2$ and repeat this process across the vehicle's trajectory as the AoH keeps changing continuously. \\
\\
\textit{\textbf{Computing AoH:}} 
Whenever a signal is received from the RRH, the AoA is determined using the MUSIC algorithm. At any given instance, if the position of the vehicle is known through prior estimates and the location of the RRH is also available, the Angle of Incoming Signal (AoIS or $\beta_k$) with respect to the reference X-axis can be calculated as follows:

\begin{equation}
\label{eq:aois}
  AoIS = \tan^{-1}(\Delta y / \Delta x)
\end{equation}

where, $\Delta y = y_{RRH} - y_{vehicle}$, $\Delta x = x_{RRH} - x_{vehicle}$, and {$\beta_k = AoA + AoH$}

Therefore, 
{
\begin{equation}
\label{eq:aoh}
 AoH (\theta_{hk}) = AoIS (\beta_k) - AoA (\theta_k)
\end{equation}}
Thus, the AoH can be derived from the above equation, representing the difference between the Angle of Incoming Signal and the Angle of Arrival.

\subsection{Communication Model} 
As discussed in Section~\ref{sec:background}, this paper explores the opportunistic use of 5G Synchronization Signal Blocks (SSBs) to estimate the vehicle's position. 
SSBs are primarily intended for network synchronization~\cite{ssb}. 
Typically, 5G base stations transmit SSBs periodically (20~ms in our case), enabling UEs to synchronize with the network. Unlike other 5G waveforms that employ Adaptive Modulation and Coding (AMC) — a feature that can introduce additional computational complexity if used for localization — SSBs operate with a fixed transmission scheme, simplifying their application in positioning tasks. We primarily use the synchronization signals for performing localization. The transmitted signal structure, path loss, and fading model are described below.

\subsubsection{Transmitted Signal Structure} 

The transmitted signal is a Cyclic-Prefix Orthogonal Frequency-Division Multiplexing (CP-OFDM) waveform as described in 3GPP 5G NR specifications~\cite{3gpp}. A sub-carrier spacing of $15$ kHz is used with CP duration $4.6 \mu \text{s}$ considering numerology 0~\cite{numer}. Given the above characteristics, we require 20 Physical Resource Blocks (PRBs) which results in a passband width $(B)$ of $3.6$ MHz $(20 \times 12 \times 15)$. Furthermore, we choose the Fast Fourier Transform (FFT) size as $256$ (with $240$ active subcarriers). We consider omnidirectional signal propagation.

\subsubsection{Path Loss and Fading Model}
The entire signal transmission takes place following the OFDM model~\cite{ofdmm} where data is split across subcarriers and sent in parallel over all subcarriers. Data is converted to a bit stream and that bit stream is parallelized and fed to the ${N}$ = $256$ point Inverse Fast-Fourier Transform (IFFT) block, $N$ being the subcarriers. CP is added, and fading is multiplied with a standard deviation of $3$ dB since we operate under $6$ GHz in the sub-$6$ GHz range. The noise model is Additive White Gaussian in nature and it incorporates thermal noise using the Boltzmann constant ($k_B = 1.38 \times 10^{-23}$ J/K) at standard noise temperature ($T_0 = 290$ K), with a typical 5G NR receiver noise figure of $5$ dB. The total noise power is calculated considering the bandwidth-dependent noise power spectral density ($N_0 = k_B T_0 B$) multiplied by the linear noise figure.
Our system consists of 1 transmitting and 4 receiving antennas. The 4 receiving antennas onboard are placed at a distance of $\lambda /2$ from each other ($\lambda$ indicating the wavelength).
Since our setting is an isolated environment, considering minimum outdoor variation, we always assume a Line of Sight (LoS) component along with Rician fading~\cite{rician}.
The path loss model follows the Friis transmission equation~\cite{friis}, where the
transmit antenna gain is $10$ dBi and received antenna gain is $5$ dBi.

\subsection{Onboard Computing System}
\label{sec:compute}
The onboard computing system executes a typical autonomous vehicle computational pipeline, comprising sensing, perception, planning, and actuation~\cite{chauffeur,avstack}. However, since this work focuses primarily on the localization module within the perception stage, we employ an OFDM wireless receiver with four antennas in a ULA as the sensing unit (shown in Figure~\ref{fig:specific}~\supc{c}) 
, along with an embedded computing unit dedicated to executing the MUSIC algorithm — both of which suffice for our use case. The sensing antennas are half-wavelength ($\lambda/2$) spaced to capture the spatial signatures of the arriving signal, modeled via a steering vector as shown in Equation~\ref{MUSIC_eq}.

For the embedded computing unit, we select the NVIDIA Jetson AGX Xavier, a multiprocessor system-on-chip (MPSoC) that is commonly used in drones, delivery vehicles, and robotics~\cite{jetson}. MPSoCs offer superior power efficiency, compact size, and reduced hardware complexity compared to alternatives such as general-purpose GPUs (GPGPUs), reconfigurable logic, and ASIC accelerators. These advantages make MPSoCs a compelling choice for practical autonomous vehicle implementations~\cite{bertozzi}.

\section{Adaptive Localization Approach}
\label{sec:loc}


While 5G RRHs transmit SSBs every 20~ms (network-dictated periodicity), vanilla MUSIC-based localization cannot match this rate due to constraints such as limited processing capacity, high power consumption, and sensitivity to dynamic noise and motion. These challenges, outlined below, motivate an adaptive localization approach:

\begin{enumerate}


\item \textbf{Processing Delays due to Hardware Limitations:} 
Although the 20~ms SSB transmission rate could theoretically support continuous localization computations, the underlying hardware may lack the necessary processing capability to operate at this frequency. Therefore, optimizing the SSB reception rate becomes essential to ensure that MUSIC computations are completed within the interval between successive SSB receptions.


\item \textbf{High Power Consumption:} 
Executing MUSIC at this frequency (50~Hz) would lead to substantial and unnecessary power consumption due to the algorithm’s computational demands. Moreover, such a high processing frequency results in an excessively fine granularity for our scenario. For example, if an ADV moves through a factory at a maximum speed of $15$ kmph, the theoretical distance margin between consecutive MUSIC computations is approximately $0.08$ m (i.e., $\text{distance} = \text{velocity} \times \text{time}$). Considering that acceptable localization granularity is typically on the order of 0.1 m or greater\footnote{For example, state-of-the-art localization methods provide accuracy in the range of 0.1~m or greater~\cite{loc1, loc2, v2x1}.}, it is wise to reduce the computation frequency to balance power consumption with localization accuracy.


\item \textbf{Environmental Noise and Motion:} 
Signal-to-Noise Ratio (SNR) conditions play a crucial role in localization accuracy, as the MUSIC algorithm depends on a clear separation of signal and noise subspaces~\cite{music_lite}. Computing MUSIC at lower SNR levels can lead to substantial error accumulation and thereby unreliable position estimation. By implementing an adaptive approach, we can adjust the computation frequency based on real-time SNR levels, enhancing the robustness of AoA estimation under fluctuating signal conditions. Furthermore, we consider cases where the vehicle may stop abruptly and resume motion, necessitating a velocity margin to trigger additional adjustments if needed.
\end{enumerate}

We overcome these challenges through the adaptive notion of
\textbf{Wake-Up Time \texttt{(WT)}}—the moment when the vehicle initiates AoA estimation via the MUSIC algorithm. This timing parameter serves as our primary control mechanism to balance accuracy and resource efficiency.
Algorithm~\ref{algo1} 
adaptively calculates \texttt{WT} based on a proportional–integral–derivative (PID) controller~\cite{pid}, and 
dynamically adjusts the \texttt{WT} by responding to two key inputs that can change at runtime: \textit{SNR error} and \textit{velocity error}. 
The \textit{SNR error} represents the deviation of the current SNR from an optimal or expected value, while the \textit{velocity error} captures discrepancies in the vehicle's estimated speed. 
The algorithm's adaptivity ensures optimal resource usage while maintaining positioning accuracy. 
\begin{algorithm}
\caption{Adaptive Wake-Up Time Calculation}
\label{algo1}
\begin{algorithmic}[1]
\Procedure{Initialize}{$t_{\text{base}},\, t_{\text{max}}$}
    \State Set $K_p$, $K_i$, $K_d$ \Comment{Proportional, integral, and derivative gains}
    \State Initialize error terms: $\epsilon_{\text{prev}} \gets 0$, $\int \epsilon\, dt \gets 0$, $\frac{d\epsilon}{dt} \gets 0$
    \State Define SNR and velocity weights: $w_{\text{snr}}$, $w_{\text{vel}}$
    \State Store $t_{\text{base}}$, $t_{\text{max}}$
\EndProcedure
\Procedure{CalculateWakeUpTime}{$\epsilon_{\text{snr}},\, \epsilon_{\text{vel}},\, \Delta t$}
    \State \textbf{Input:} $\epsilon_{\text{snr}}$ (normalized SNR error), $\epsilon_{\text{vel}}$ (normalized velocity error), $\Delta t$ (time interval)
    \State \textbf{Output:} Adjusted wake-up time \texttt{(WT)}
    \State $\epsilon_{\text{combined}} \gets w_{\text{snr}} \cdot \epsilon_{\text{snr}} + w_{\text{vel}} \cdot \epsilon_{\text{vel}}$
    \State $\int \epsilon\, dt \gets \int \epsilon\, dt + \epsilon_{\text{combined}} \cdot \Delta t$
    \State $\frac{d\epsilon}{dt} \gets \frac{\epsilon_{\text{combined}} - \epsilon_{\text{prev}}}{\Delta t}$
    \State $u \gets K_p \cdot \epsilon_{\text{combined}} + K_i \cdot \int \epsilon\, dt + K_d \cdot \frac{d\epsilon}{dt}$
    \State $\epsilon_{\text{prev}} \gets \epsilon_{\text{combined}}$
    \State $WT \gets t_{\text{base}} + u$
    \State $WT \gets \min\big(\max(WT, t_{\text{base}}), t_{\text{max}}\big)$
    \State \Return \texttt{WT}
\EndProcedure
\end{algorithmic}
\end{algorithm}

Algorithm~\ref{algo1} comprises 5 main functional components:
\begin{itemize}
\item \textbf{Initialization \texttt{(Lines 1-6)}:} \texttt{Line 2} sets the PID gains ($K_p$, $K_i$, $K_d$) which control how aggressively the system responds to errors; \texttt{line 3} initializes error terms; \texttt{line 4} defines weights that balance the importance of signal quality versus motion stability; and \texttt{line 5} stores time bounds that establish the operational limits. This phase creates a framework for adaptive decision-making that can prioritize either responsiveness or stability.

\item \textbf{Input and Output  \texttt{(Lines 7-9)}:} \texttt{Line 8} specifies inputs that capture both environmental quality (SNR) and motion characteristics (velocity), enabling context-aware adaptation; \texttt{line 9} defines the output as an adaptive wake-up time. This multi-factor approach allows the system to respond intelligently to complex, changing conditions by adjusting measurement frequency.

\item \textbf{Error Processing \texttt{(Lines 10-12)}:} \texttt{Line 10} fuses individual errors through a weighted sum, reflecting the system's designed sensitivity to different error sources; \texttt{line 11} updates the integral term to capture persistent conditions over time; \texttt{line 12} computes the derivative term to detect rapid changes. Together, these operations provide temporal awareness across multiple timescales, addressing both gradual drifts and sudden environmental shifts.

\item \textbf{PID Calculation  \texttt{(Lines 13-14)}:} \texttt{Line 13} computes a corrective adjustment using classical control theory that balances immediate response, historical patterns, and trend prediction; \texttt{line 14} updates the error history for continuity. This approach prevents oscillatory behavior while maintaining responsiveness, a critical balance in adaptive scheduling applications.

\item \textbf{Result Generation  \texttt{(Lines 15-17)}:} \texttt{Line 15} calculates the new wake-up time by applying the correction to the base interval; \texttt{line 16} enforces boundary constraints to prevent extreme scheduling decisions; \texttt{line 17} delivers the optimized timing. The bounded result ensures the system operates within defined constraints, adaptively maximizing information quality by waking up more frequently under favorable conditions (high SNR, stable velocity) and conserving resources by reducing unnecessary measurements during unfavorable, high-error scenarios—effectively balancing efficiency and accuracy.
\end{itemize}
\begin{figure*}
    \centering
    \includegraphics[width=0.75\textwidth]{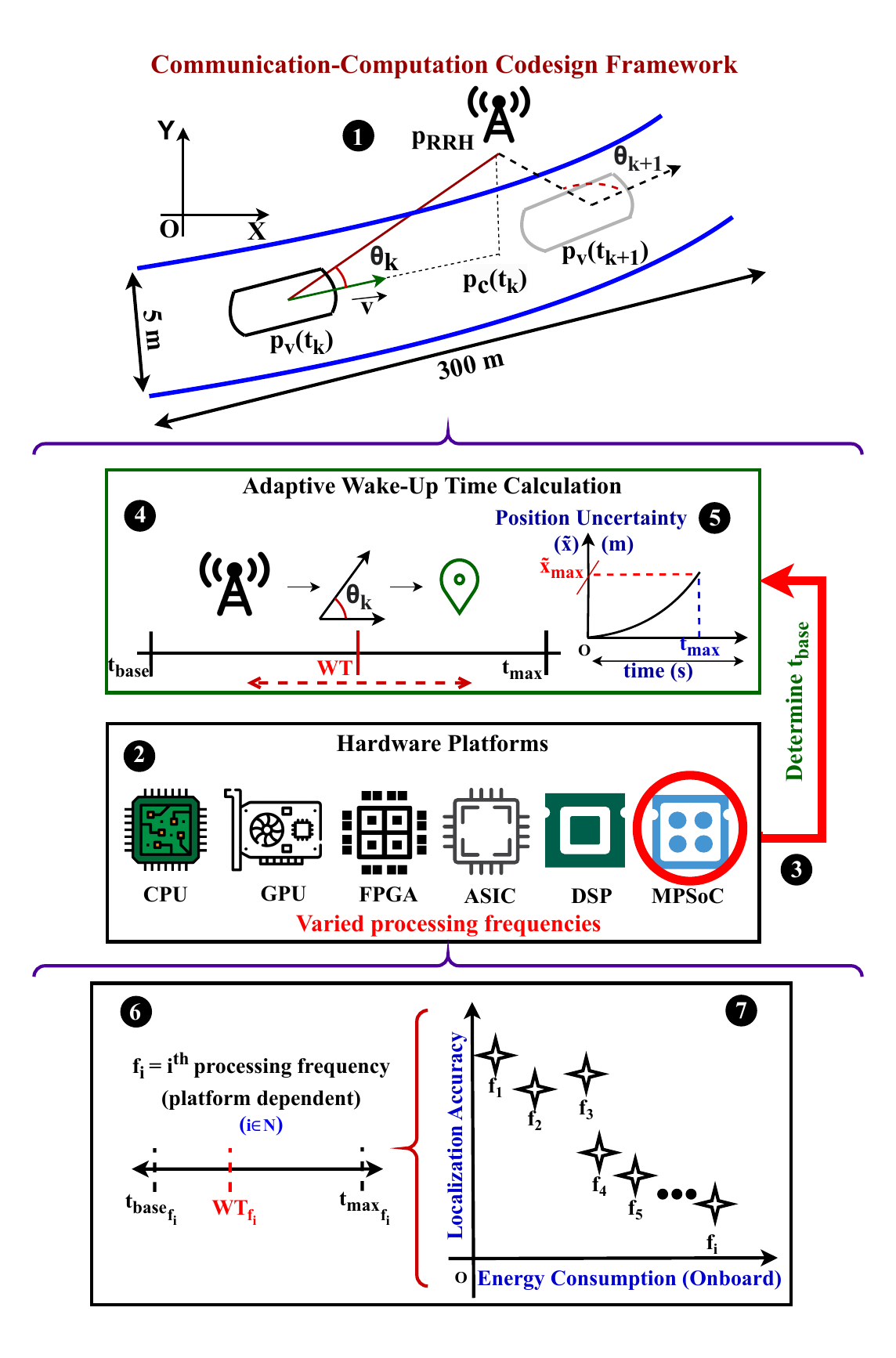}
    \caption{Communication-Computation Codesign Framework of ACCESS-AV. We select an MPSoC (NVIDIA Jetson AGX Xavier) as our hardware platform which is circled in red (\circlednum{2}). 
    }
    \label{fig:acc2}
\end{figure*}

Through this adaptive \texttt{WT} approach, our localization system effectively balances computational efficiency, power consumption, and positioning accuracy across varying environmental and motion conditions.

\medskip
{\noindent\textbf{\emph{Robustness via Adaptation.}}  
Classical MUSIC, while offering high-resolution estimates, is known to degrade under low Signal-to-Noise Ratio (SNR) and during sudden changes in vehicle motion. Rather than modifying the MUSIC algorithm itself — which is not the focus of this work — we introduce a \emph{system-level adaptation mechanism} that enhances localization robustness by monitoring real-time SNR and velocity conditions. Under unfavorable conditions, the controller adaptively defers AoA estimation, thereby avoiding unreliable computations and reducing the risk of cumulative localization errors.  
As shown in Section~\ref{sec:eval}, this mechanism enables \emph{ACCESS-AV} to maintain worst-case localization errors below 0.3 meters across all power modes, while saving up to 43.09\% energy onboard and outperforming fixed-time or vanilla MUSIC, ESPRIT, and Root-MUSIC baselines. This adaptive strategy serves as a robust and efficient localization solution, well-suited for deployment on resource-constrained platforms in smart factory environments.}

\section{Communication-Computation Codesign Framework}
\label{sec:codesign}


Figure~\ref{fig:acc2}
shows \emph{\papername{}}'s communication-computation codesign framework. 
This framework allows users to explore the accuracy-energy trade-offs based on set quality constraints for different hardware frequency settings
in diverse operating conditions.
As discussed in Section~\ref{sec:loc}, although the vehicle can theoretically be localized every 20~ms ($t_\text{theoretical}$), practical challenges such as limited processing capacity, high power consumption, and sensitivity to dynamic noise and motion prevent this from being feasible.
Therefore, based on the localization task at hand~\supc{1}, an appropriate hardware platform is selected initially from those in Figure~\ref{fig:acc2}.
In this study, \emph{\papername{}} operates with an MPSoC (the  NVIDIA Jetson AGX Xavier) as its hardware platform~\supc{2}. 
As mentioned in Section~\ref{sec:compute}, MPSoCs offer superior power efficiency, compact size, and reduced hardware complexity~\cite{bertozzi}, in addition to executing end-to-end autonomous vehicle computational pipelines at scale~\cite{chauffeur, crx}. 

Once a platform is selected, its processing frequency determines the corresponding base wake-up time ($t_\text{base}$)~\supc{3}, ensuring that MUSIC computations are completed faster than the SSB reception rate. 
We also call this the hardware-constrained baseline (\(t_{\text{base}}\)), which is greater than or equal to $t_\text{theoretical}$.
Given $t_\text{base}$, we then determine a suitable $t_\text{max}$ to establish the interval \([t_{\text{base}}, t_{\text{max}}]\)~\supc{4}, within which the vehicle determines its \texttt{WT} for AoA estimation and thereby localizes itself. $t_\text{max}$ is user-specific and must ensure that \texttt{WT} is determined before the vehicle exceeds the position uncertainty upper bound \(\tilde{x}_{\max}\)~\supc{5}, beyond which MUSIC computations may not provide reliable localization. 
\texttt{WT} determination is influenced not only by the platform's processing frequency but also by environmental factors like SNR as mentioned in Section~\ref{sec:loc}. 

Once the interval \([t_{\text{base}}, t_{\text{max}}]\) is determined, \texttt{WT} is calculated using Algorithm~\ref{algo1}. 
Since most platforms today (e.g., NVIDIA Jetson AGX Xavier) support
multiple processing frequencies, each frequency results in a different \(t_{\text{base}}\), leading to distinct \([t_{\text{base}}, t_{\text{max}}]\) intervals. 
This results in multiple intervals \([t_{\text{base}_{f_i}}, t_{\text{max}_{f_i}}\)], where \(f_{i}\) corresponds to the $i^{th}$ available processing frequency of the platform, with $i \in \mathbb{N}$~\supc{6}. For each \(f_{i}\), we perform localization, evaluate accuracy across the X and Y directions, and measure corresponding onboard energy consumption. These results help construct a design space~\supc{7}, allowing us to extract optimal design points balancing accuracy and energy needs under user-defined quality constraints.

\section{Experimental Setup}
\label{sec:expt}
\begin{figure*}
    \centering
    \includegraphics[width=1\textwidth]{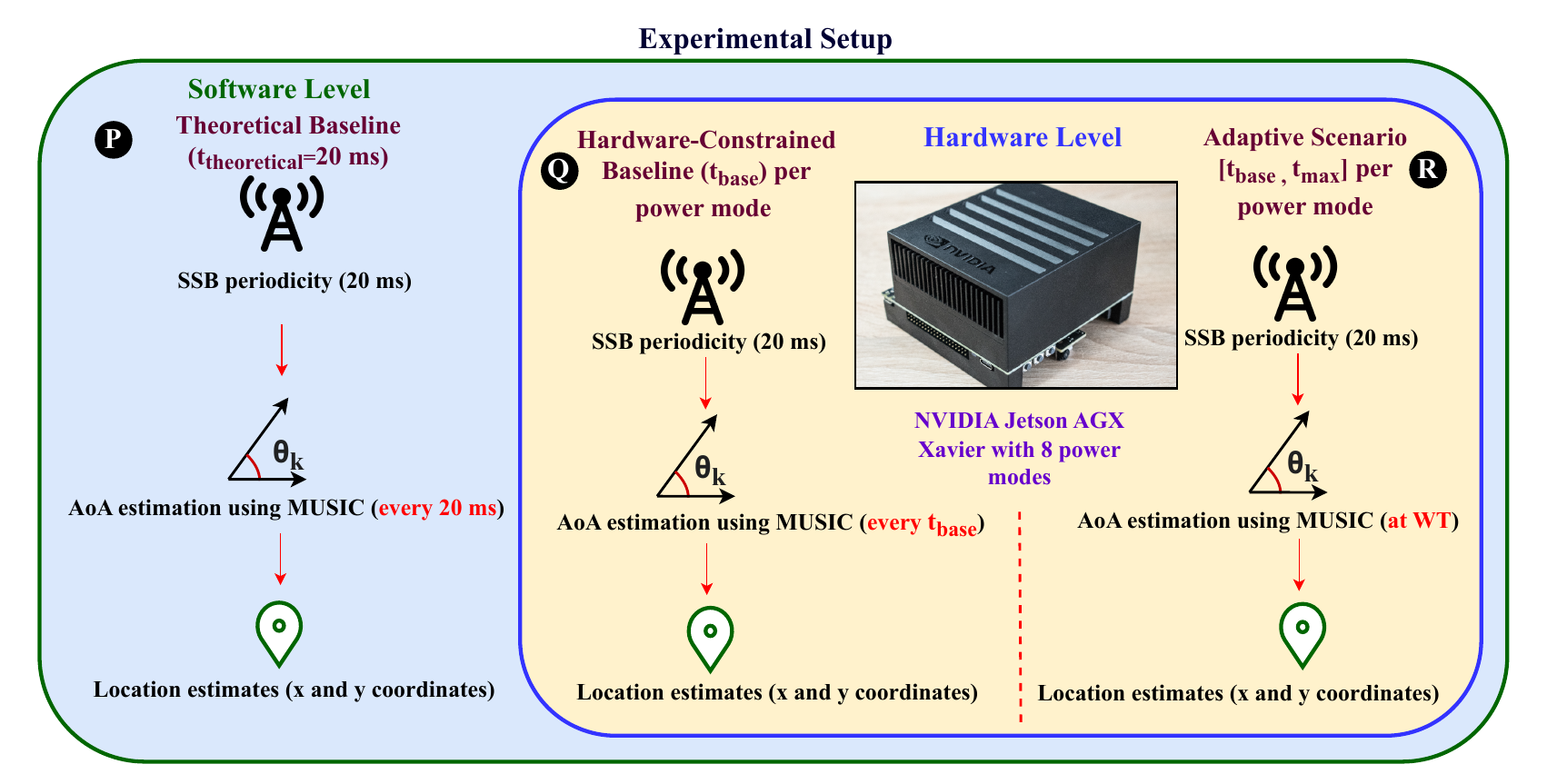}
    \caption{Experimental Setup of \papername{}. At the software level, all cases, i.e.,~\circlednum{P},~\circlednum{Q}, and~\circlednum{R} are considered. At the hardware level, only~\circlednum{Q}, and~\circlednum{R} are considered owing to hardware constraints.}
    \label{fig:expt}
\end{figure*}
Our experimental setup is shown in Figure~\ref{fig:expt}.
To assess accuracy, we simulate the system in Python at the software level. We then conduct an energy consumption study at the hardware level using the NVIDIA Jetson AGX Xavier (32 GB) board~\cite{jetson}, 
testing eight power modes~\cite{power_modes} with varied CPU frequency levels (for vanilla MUSIC, ESPRIT, Root-MUSIC, and adaptive MUSIC), 
listed
as \emph{\texttt{MODE 0 -- MODE 7}} in
Table~\ref{tab:power_modes}. 
For each CPU frequency, we dynamically tune the corresponding \([t_\text{base}, t_\text{max}]\) interval in case of the adaptive scenario,
allowing us to analyze the impact of varying time intervals on localization accuracy.
\begin{table}[htbp]
\centering
\caption{CPU frequencies per power mode for NVIDIA Jetson AGX Xavier~\cite{power_modes} and corresponding base wake-up times for MUSIC, ESPRIT, and Root-MUSIC.}
\label{tab:power_modes}
\setlength{\tabcolsep}{3pt} 
\begin{tabular}{|c|c|c|c|c|}
\hline
\textbf{\shortstack{Power Mode}} & \textbf{CPU freq. (MHz)} & \textbf{\shortstack{Base \\ Wake-Up \\ Time (MUSIC)}} & \textbf{\shortstack{Base \\ Wake-Up \\ Time (ESPRIT)}} & \textbf{\shortstack{Base \\ Wake-Up \\ Time (Root-MUSIC)}} \\ \hline
MODE 0 & 2265.6 & 80~ms & 65~ms & 70~ms \\ \hline
MODE 1 & 1200 & 150~ms & 130~ms & 135~ms \\ \hline
MODE 2 & 1200 & 150~ms & 130~ms & 135~ms \\ \hline
MODE 3 & 1200 & 150~ms & 130~ms & 135~ms \\ \hline
MODE 4 & 1450 & 120~ms & 110~ms & 110~ms \\ \hline
MODE 5 & 1780 & 100~ms & 90~ms & 95~ms \\ \hline
MODE 6 & 2100 & 100~ms & 70~ms & 75~ms \\ \hline
MODE 7 & 2188 & 100~ms & 70~ms & 75~ms \\ \hline
\end{tabular}
\end{table}


The AoA algorithms considered (vanilla MUSIC, ESPRIT, Root-MUSIC, and adaptive MUSIC) employ a configuration of four antennas with 20 snapshots. To evaluate our approach, we consider three distinct scenarios:
\begin{enumerate}
    \item A theoretical baseline ($t_\text{theoretical}$) of 20 ms, corresponding to the SSB periodicity-       Figure~\ref{fig:expt}~\supc{P}
    \item A hardware-constrained baseline ($t_\text{base}$) per power mode including each AoA algorithm considered (vanilla MUSIC, ESPRIT, Root-MUSIC), for benchmarking, as shown in Table~\ref{tab:power_modes} 
    - Figure~\ref{fig:expt}~\supc{Q}
    \item An adaptive scenario of MUSIC using our~\emph{\papername{}} approach where the lower limit is determined by the hardware-constrained baseline of MUSIC($t_\text{base}$) per power mode, and the upper limit ($t_\text{max}$) is set at 240 ms- Figure~\ref{fig:expt}~\supc{R}
\end{enumerate}

The upper limit ($t_\text{max}$) of 240 ms is determined using the kinematic relation, ${distance} = {velocity} \times {time}$. Given that the vehicle travels at a uniform speed of 15 kmph (4.17 mps) for most of its path, it covers 1 m in approximately 240 ms. As mentioned earlier, this upper limit is user-specific and can be adjusted based on quality constraints. For instance, if a user is comfortable with a maximum position uncertainty (\(\tilde{x}_{\max}\)) of 2 m, the upper limit would be approximately 480 ms.
For our upper limit of 240 ms, we evaluate the system by running it 100 times and averaging the results.

For the energy consumption study, we run the hardware-constrained baselines of MUSIC, ESPRIT, and Root-MUSIC (Figure~\ref{fig:expt}~\supc{Q}) along with the adaptive (Figure~\ref{fig:expt}~\supc{R}) case of MUSIC for each power mode, measuring total energy consumption over the vehicle's entire trajectory. The adaptive case ensures \texttt{WT} selection remains within the hardware-constrained baseline of MUSIC ($t_\text{base}$) and the user-defined theoretical distance margin (\(\tilde{x}_{\max}\)) of 1 m, with an upper bound ($t_\text{max}$) of 240 ms.

For the PID controller responsible for adaptive \texttt{WT} calculation, the PID gains are set to 0.75 (Proportional Gain), 0.08 (Integral Gain), and 0.195 (Derivative Gain), based on iterative tuning and empirical testing. The SNR and velocity weights are set to 0.6 and 0.4, respectively.

\section{Evaluation}
\label{sec:eval}

We perform 4 sets of evaluations to demonstrate the efficacy of \emph{\papername{}}: (1) accuracy analysis at the software level, (2) energy consumption study at the hardware level, (3) design space exploration for analysing accuracy vs. energy consumption trends, and (4) cost estimation. 
\subsection{Accuracy Analysis}
\begin{figure}
\centering
\includegraphics[width=0.9\textwidth]{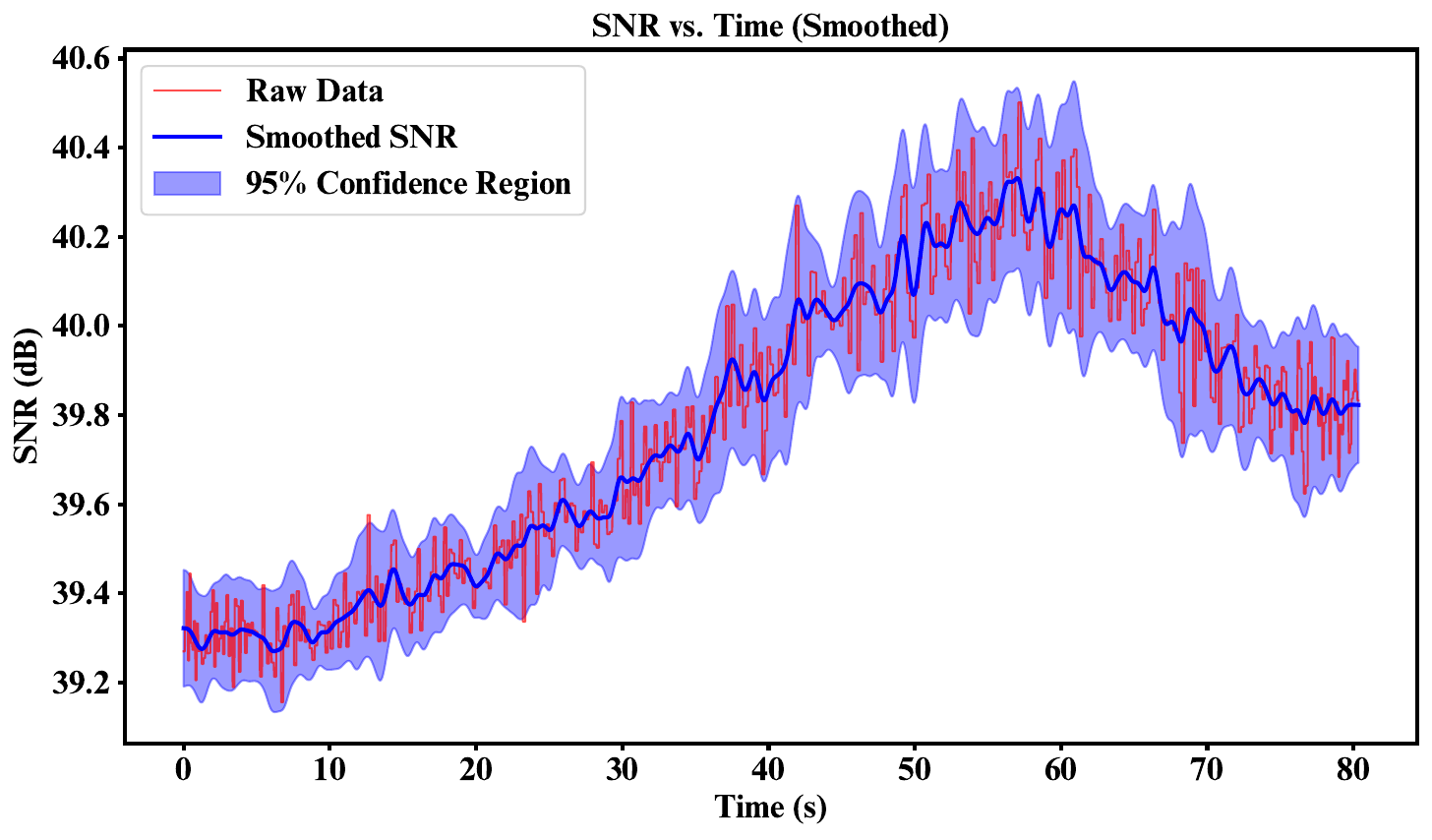}
\caption{Signal-to-Noise Ratio (SNR) trend across the vehicle's trajectory over time.}
\label{fig:snr}
\end{figure}
\begin{figure*}
    \centering
    \includegraphics[width=0.92\textwidth]{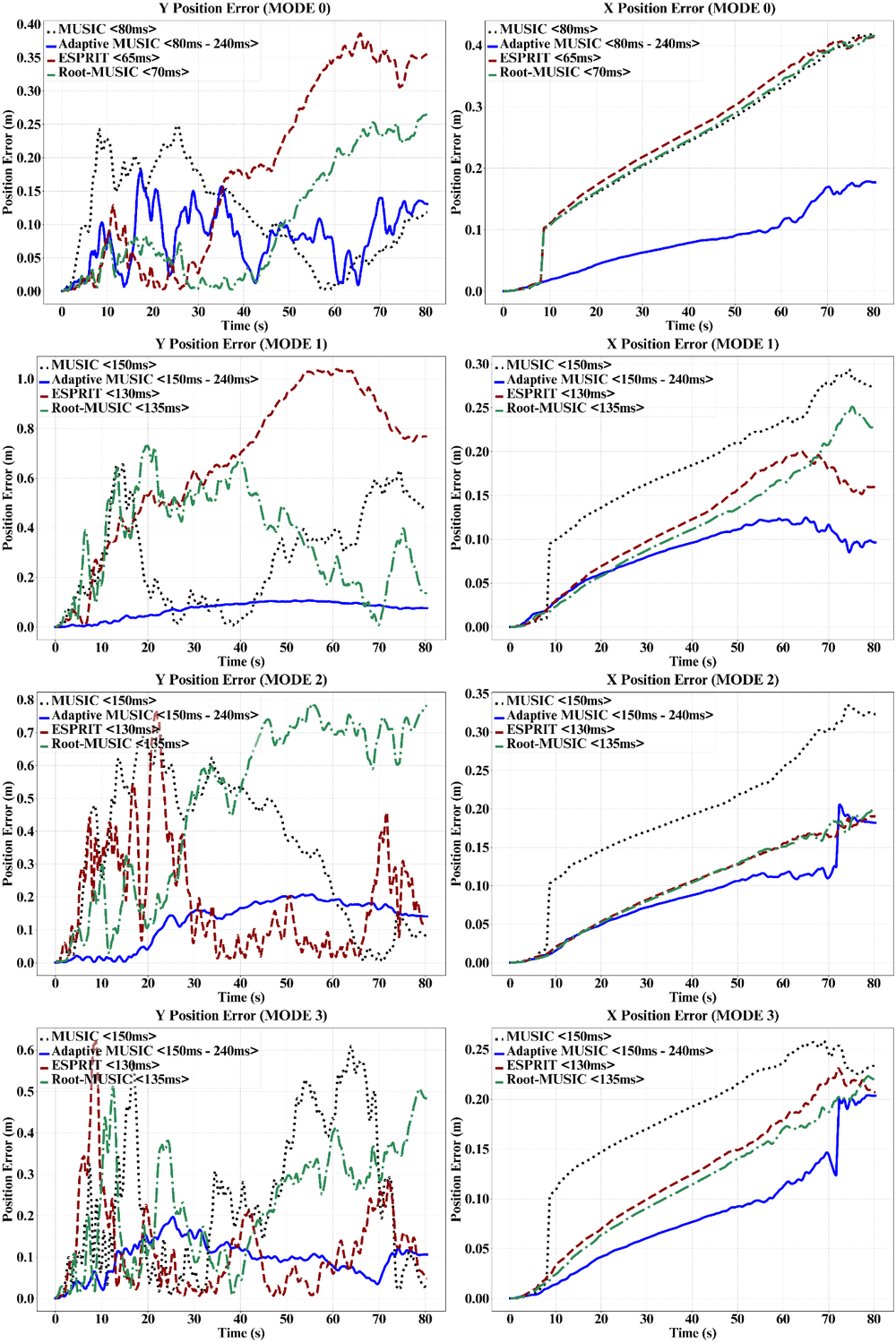}
    \caption{Accuracy Analysis (across both Y and X directions) for MUSIC (in black), ESPRIT (in red), Root-MUSIC (in green) and adaptive MUSIC scenario (in blue) across power modes 0 to 3.}
    \label{fig:acc}
\end{figure*}
\begin{figure*}
    \centering
    \includegraphics[width=0.92\textwidth]{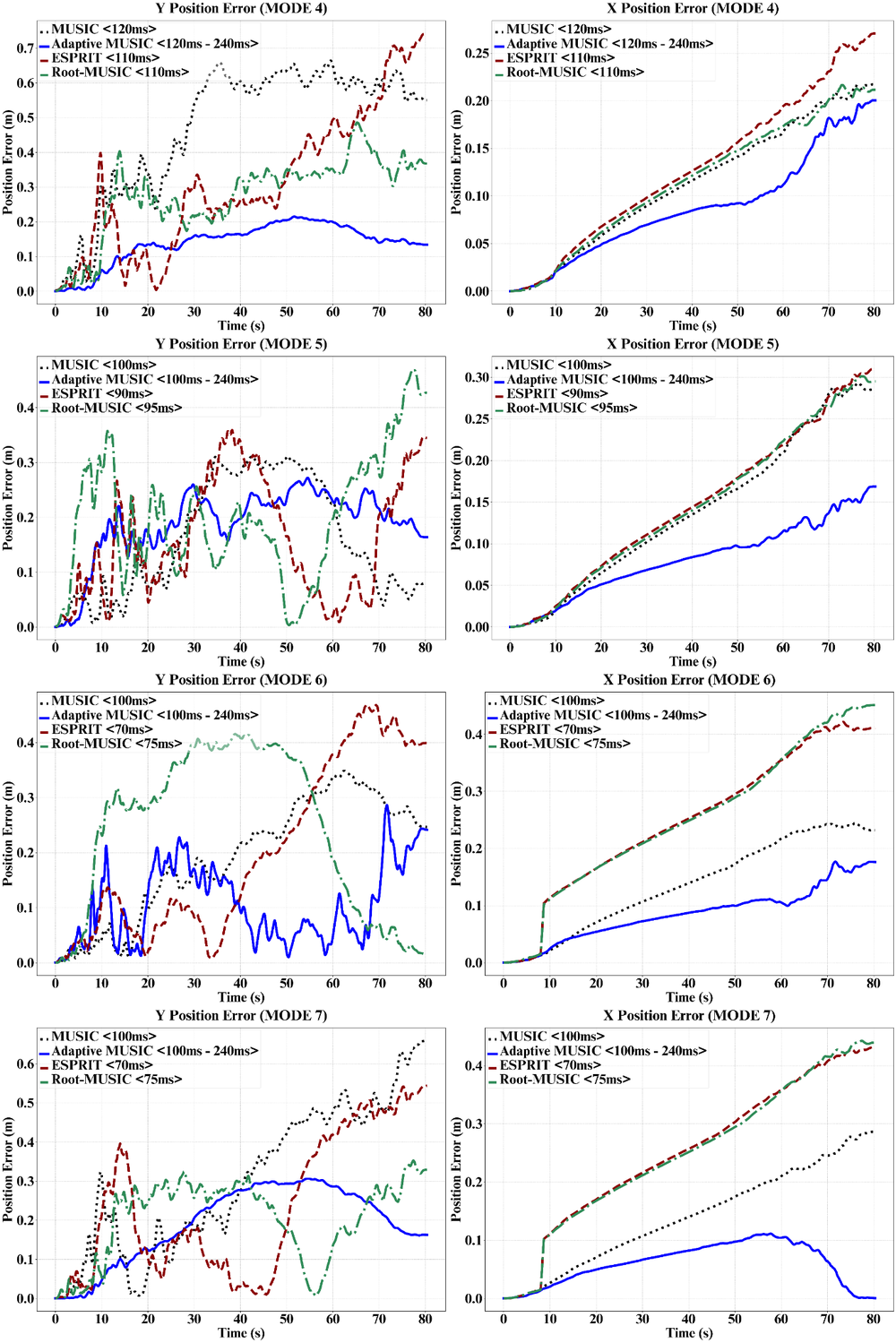}
    \caption{Accuracy Analysis (across both Y and X directions) for MUSIC (in black), ESPRIT (in red), Root-MUSIC (in green) and adaptive MUSIC scenario (in blue) across power modes 4 to 7.}
    \label{fig:acc_2}
\end{figure*}

To evaluate the accuracy of our proposed approach, we first profile the SNR along the ADV's trajectory as shown in Figure~\ref{fig:snr}, given that the accuracy of the MUSIC algorithm is inherently dependent on SNR, as discussed earlier. Our analysis reveals that SNR increases as the vehicle approaches the RRH, reaching its peak when in closest proximity before gradually declining. In practical deployments, however, this decline is mitigated through effective Radio Frequency (RF) planning, ensuring that a new RRH is detected along the vehicle's trajectory. 

We perform accuracy analysis for three scenarios: (1) a theoretical baseline of 20 ms, (2) hardware-constrained baselines of MUSIC, i.e., 80 ms, 100 ms, 120 ms, and 150 ms corresponding to specific power modes as shown in Table~\ref{tab:power_modes}, and (3) an adaptive scenario with an upper limit of 240 ms. Figure~\ref{fig:acc} presents the corresponding accuracy levels for the three scenarios in both the X and Y directions of the vehicle's trajectory.  Also, we benchmark our adaptive approach against other state-of-the-art AoA algorithms, namely, ESPRIT and Root-MUSIC, that have their own respective hardware-constrained baselines as shown in Table~\ref{tab:power_modes}.

For MUSIC's hardware-constrained baselines, based on power modes as detailed in Table~\ref{tab:power_modes}, we see that \texttt{MODE 0} has a baseline of 80 ms; \texttt{MODES 1, 2, and 3} have a baseline of 150 ms; \texttt{MODE 4} has a baseline of 120 ms; \texttt{MODES 5, 6, and 7} have a baseline of 100 ms. For the case of theoretical baseline ($t_\text{theoretical}$= 20 ms), we see that the worst-case error is 0.8~m on average across vanilla MUSIC, ESPRIT, and Root-MUSIC.

Hence, from Figures~\ref{fig:acc} and~\ref{fig:acc_2}, we make 6 major observations. First, the adaptive approach consistently maintains position errors within a 0.3~m boundary in both the X and Y directions across all power modes, demonstrating reliable localization accuracy using 5G SSBs. Second, it significantly outperforms the hardware-constrained baselines of all AoA algorithms considered throughout most of the vehicle's trajectory, irrespective of the power mode configuration. Third, the hardware-constrained baselines, which rely on fixed wake-up times, fail to account for external factors such as SNR variations, resulting in cumulative errors that can lead to catastrophic localization failures as the vehicle moves. By dynamically adjusting wake-up times based on environmental conditions, the adaptive approach effectively mitigates these risks. Fourth, we see that among all AoA algorithms, ESPRIT performs worst in terms of accuracy, reaching worst-case error of over 1~m in the Y-direction. This is because ESPRIT, having the lowest base wake-up time per power mode (shown in Table~\ref{tab:power_modes}), fails to account for SNR and velocity errors, which results in higher localization error accumulation due to shorter wake-ups. Fifth, we see that even Root-MUSIC is unable to maintain an error margin of 0.3~m, and is having a worst-case position error of 0.8~m in the Y-direction, thereby supporting the need for an adaptive approach. Sixth, we see that  \texttt{MODES 1, 2, and 3}, despite having the largest wake-up time for MUSIC among all modes (150~ms), achieve extremely low
position errors of less than 0.2~m in both the X-direction and Y-directions, making them promising candidates for localization requirements. Overall, it is important to note that our adaptive \emph{ACCESS-AV} consistently limits the maximum position error to within 0.3~m across all cases, making it a highly suitable method for real-time localization scenarios while saving onboard energy.

\subsection{Energy Consumption Study}

\begin{figure}
\centering
\includegraphics[width=0.95\textwidth]{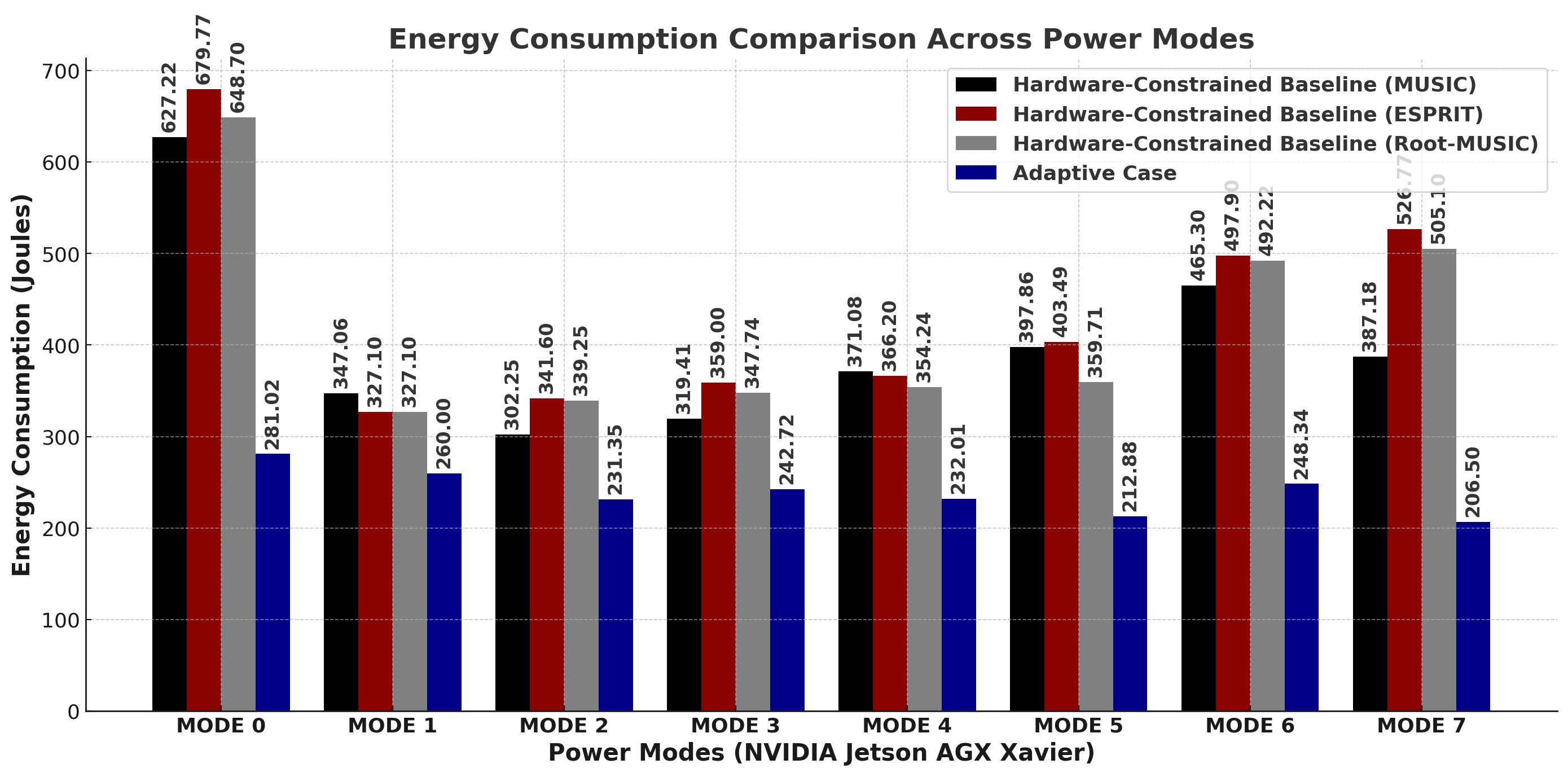}
\caption{Energy consumption (in Joules) for the hardware-constrained baselines of MUSIC (in black), ESPRIT (in red), and Root-MUSIC (in gray), and the adaptive MUSIC case (in blue), per power mode across the vehicle's trajectory.}
\label{fig:hw}
\end{figure}
To evaluate system-level implications and quantify energy efficiency gains of \emph{ACCESS-AV}, we conduct an energy consumption study on the NVIDIA Jetson AGX Xavier platform and benchmark our approach against vanilla MUSIC, ESPRIT, and Root-MUSIC. As shown in Figure~\ref{fig:hw}, the energy consumption measurements (in Joules) across various CPU frequency levels compare the hardware-constrained baselines to the adaptive approach.

We make 4 major observations from Figure~\ref{fig:hw}. 
First, we see that across all power modes, the adaptive case is able 
to save significant energy compared to respective baselines. Second, we see that all AoA algorithms consume the most amount of energy in \texttt{MODE 0}, 627.22 J for vanilla MUSIC, 679.77 J for ESPRIT, and 648.70 J for Root-MUSIC. This is not surprising because \texttt{MODE 0} has a CPU frequency of 2265.6 MHz which is the highest among all modes, and all algorithms have their lowest wake-up times in this mode, i.e., 80 ms for vanilla MUSIC, 65 ms for ESPRIT, and 70 ms for Root-MUSIC. This means that the MUSIC algorithm is executed over 1000 times during the vehicle's 80.33-second trajectory, ESPRIT is executed over 1235 times, and Root-MUSIC is executed over 1147 times. Since these algorithms are CPU-intensive~\cite{music_lite}, such high computation frequency leads to substantial energy consumption. Third, we see that for \texttt{MODES 6 and 7}, energy gains are above 52\% using adaptivity. Similar reasoning applies here: these modes have baseline computation intervals of 100~ms (MUSIC), 70~ms (ESPRIT), and 75~ms (Root-MUSIC), meaning MUSIC is computed over 800 times during the vehicle's 80.33-second trajectory, ESPRIT is computed over 1147 times, and Root-MUSIC is computed over 1071 times. Adaptivity in these modes reduces energy consumption effectively, as enabled by \emph{ACCESS-AV}. Furthermore, these modes feature high CPU frequencies, second only to \texttt{MODE 0}, which explains the high baseline energy consumption. Finally, we see that our solution scales significantly towards saving energy onboard, for example, if we consider operating in \texttt{MODE 0} towards performing localization, one vehicle over a 24-hour runtime saves 398.84 kJ using \emph{\papername{}}. Scaling this to a fleet of 100 vehicles, \emph{\papername{}} helps save 39.884 MJ cumulatively, which is roughly equivalent to the energy consumed by two refrigerators per day~\cite{fridge}. 

\subsection{Design Space Exploration}
\begin{figure}
\centering
\includegraphics[width=0.75\textwidth]{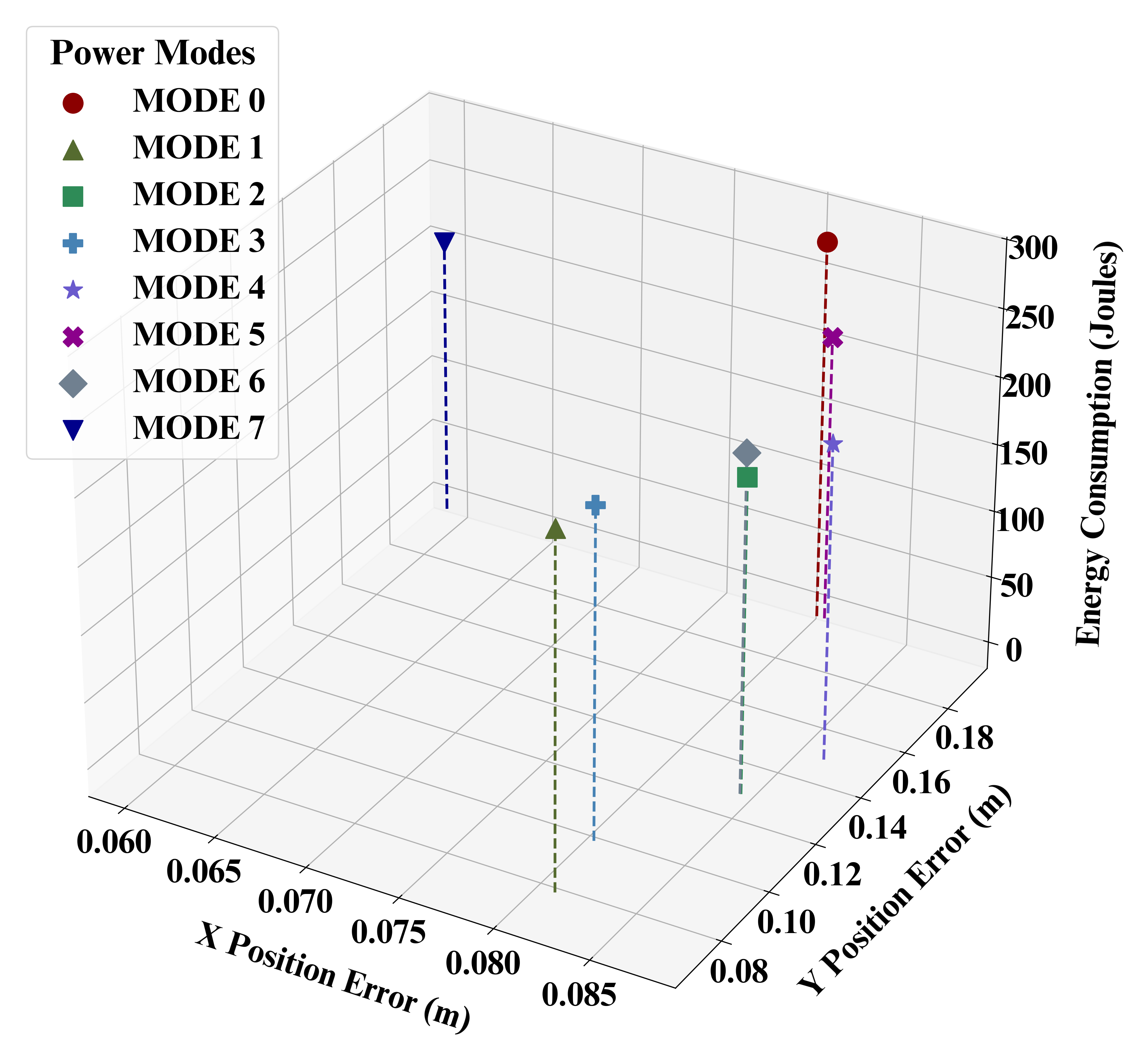}
\caption{Design space exploration for accuracy vs. energy consumption analysis.}
\label{fig:dse}
\vspace{-3ex}
\end{figure}

Having determined localization accuracy and onboard energy consumption for each power mode, the next step is to extract optimal design points that meet user-defined quality constraints. These constraints may involve a specific energy budget, a target accuracy level, or a balance between both to identify the most suitable power mode. To visualize this trade-off, we present Figure~\ref{fig:dse}, a 3D grid that illustrates the performance of various power modes of the NVIDIA Jetson AGX Xavier while executing our localization approach.
The three axes represent position errors in the X and Y directions (in meters) and onboard energy consumption for each power mode. All results correspond to the adaptive case only. By analyzing this design space, we can draw several key observations. For instance, given an energy budget of 210 J, \texttt{MODE 7} emerges as the most efficient choice, consuming only 206.5 J while providing an RMS error below 0.2 m. In another case, if we have an RMS error tolerance margin of 0.15 m coupled with an energy budget below 260 J, \texttt{MODES 2, 3, and 6} satisfy our criteria. Therefore, based on user-defined quality constraints—whether prioritizing accuracy, energy efficiency, or both—multiple design points can be extracted.

\subsection{Cost Estimation}
We now present cost estimation  
to highlight \emph{ACCESS-AV}'s  role in promoting sustainable development by minimizing resource usage and system complexity. 

First, at the factory infrastructure level, \emph{ACCESS-AV} leverages RRHs for signaling instead of dedicated RSUs, resulting in significant cost savings. According to Tonguz~\textit{et al.}~\cite{rsucost}, the capital cost of an RSU ranges from \$13,000 to \$15,000 per unit, with an additional annual operation and maintenance expense of up to \$2,400 per unit. These costs, reported in 2013, have only escalated by 2025 keeping in mind inflation rates, representing a substantial financial burden for businesses. For a factory deploying multiple RSUs, the expenses scale significantly. For instance, maintaining four RSUs would require an initial investment of \$52,000 to \$60,000, along with an annual operational cost of approximately \$9,600. Such expenditures impose a considerable financial strain on business owners. Also, deploying multiple RSUs results in an increase in both embodied and operational carbon-footprint~\cite{sze,jones}. By contrast,~\emph{ACCESS-AV} leverages the factory’s existing 5G infrastructure, eliminating the need for additional RSUs and thereby offering a cost-effective, carbon-efficient and scalable alternative.

Second, \textit{ACCESS-AV} offers substantial cost savings by minimizing onboard sensor requirements. Unlike conventional localization systems that rely on multiple expensive sensors such as LiDAR, high-end cameras and radar, \textit{ACCESS-AV} employs only an OFDM wireless receiver, significantly reducing costs for business owners. For example, a single Velodyne VLP-16 LiDAR sensor costs approximately \$4,000~\cite{velodyne}, while a 5G network adapter with an embedded wireless receiver costs just \$30.71~\cite{amazon} and can be easily plugged into the NVIDIA Jetson platform—representing over a $130\times$ cost reduction compared to LiDAR sensors. Traditional localization methods often incorporate sensor fusion techniques that integrate data from multiple heterogeneous sensors~\cite{fuse}. While such approaches enhance localization accuracy, they introduce additional system complexity and computational overhead~\cite{complex}. In contrast, \textit{ACCESS-AV} adopts a lightweight strategy by leveraging 5G SSBs for localization, requiring only an OFDM wireless receiver onboard. This not only simplifies system architecture but also reduces operational complexity.

Moreover, minimizing the number of onboard sensors increases the available payload capacity for ADVs, both in terms of material storage and energy efficiency. Additionally, since \textit{ACCESS-AV} employs an adaptive localization approach, it maintains an accuracy threshold of 0.3 m while saving 43.09\% energy onboard. Therefore, by reducing system complexity and energy demands, \textit{ACCESS-AV} enhances overall efficiency, making it a more sustainable solution for large-scale deployment. Overall, \textit{ACCESS-AV} provides a scalable solution towards energy-efficient and sustainable localization in smart factories while satisfying user-defined quality constraints.
\section{Related Works}
\label{sec:related}
Autonomous vehicle localization can be broadly categorized into three primary approaches: conventional methods, machine learning-based techniques, and V2X-enabled solutions~\cite{localization_survey}. 

Conventional methods often rely on multi-sensor fusion to achieve high precision. For instance, Wan \textit{et al.}~\cite{loc1} proposed a robust system that combines LiDAR, GNSS, and IMU data through an error-state Kalman filter with uncertainty estimation, achieving centimeter-level accuracy (5–10 cm RMS) in diverse urban environments. While their approach demonstrates high precision, the study does not explicitly address the inherent system complexity and computational overhead associated with real-time multi-sensor integration. Furthermore, it does not analyze scalability challenges in resource-constrained scenarios. Similarly, Bauer \textit{et al.}~\cite{loc2} introduced a particle filter-based framework that integrates HD maps with GNSS, IMU, and odometry data, achieving a best-case RMSE of 1.2 m. However, their work does not evaluate the computational burden or practical limitations of deploying such a system in dynamic, large-scale environments, leaving gaps in understanding its real-world feasibility.

Machine learning-based techniques have also been explored for localization. Cavalcante \textit{et al.}~\cite{dnn1} developed a deep neural network-based global visual localization system (DeepVGL) that is integrated within an autonomous driving framework. This system leverages multiple sensors—such as cameras, LiDAR, GNSS, and IMU—for tasks like occupancy grid mapping and precise pose tracking. DeepVGL achieves 96\% localization accuracy within 5 m; however, the approach relies on extensive training datasets and demands significant computational resources during inference, which poses challenges for deployment on embedded systems. Moreover, while the broader framework benefits from multi-sensor fusion, the authors do not fully analyze the complexities—such as increased latency and energy consumption—that arise from integrating and processing data from multiple sensors. These limitations underscore the need for lightweight architectures and efficient data pipelines to balance accuracy with operational constraints. In another work, Zhang \textit{et al.}~\cite{dnn2} developed a neural network using attention mechanisms and shared MLPs to encode 3D semantic features (e.g., lane lines and traffic signs) from camera and wheel speed sensor data, eliminating manual data association. The model integrates camera, IMU, and GNSS via an invariant extended Kalman filter (InEKF). However, it requires fixed-dimensional tensor inputs, relies on synthetic training data and HD maps, and does not address challenges in dynamic environments or computational overhead.

V2X-enabled solutions offer another promising avenue for localization. Huang \textit{et al.}~\cite{v2x1} proposed a reconfigurable intelligent surface (RIS)-assisted near-field localization method using received signal strength measurements. Their framework employs weighted least squares (WLS) and alternate iteration to estimate positions, achieving RMSEs of 0.36 m (range) and 0.82° (azimuth) at 4 dB noise levels. However, the method assumes predefined RIS subpart divisions and static environments, leaving scalability in dynamic scenarios and the computational costs of real-time phase adjustments unexplored. In another work, Zhu \textit{et al.}~\cite{zhu} addressed the localization of autonomous vehicles in tunnels using a V2I cooperative methodology. Their system employs a roadside multi-sensor fusion framework to detect vehicle positions and transmit them to the vehicle-side subsystem for improved localization. While this approach enhances accuracy by mitigating GNSS limitations in tunnels, it introduces significant computational overhead due to the real-time processing of LiDAR, camera, and radar data. Additionally, deploying and maintaining dense roadside units equipped with LiDAR and edge computing poses scalability challenges.

Therefore, across all the localization approaches discussed above, a common challenge emerges: the insufficient consideration of system-level complexity, resource constraints, and budget limitations-- both onboard and beyond. This gap underscores the need for a framework like \emph{\papername{}} that delivers a lightweight localization solution while minimizing resource usage. In scenarios such as smart factories—where extensive network and automation infrastructure is already in place—we believe our method presents a promising energy- and resource-efficient solution  with sustainability at its core.

\section{Conclusion and Future Work}
\label{sec:conclusion}

We introduced \emph{\papername{}}, an innovative adaptive Vehicle-to-Infrastructure (V2I)-based localization framework specifically tailored for Autonomous Delivery Vehicles (ADVs) operating in smart factories. By intelligently leveraging existing 5G Synchronization Signal Blocks (SSBs) transmitted by Remote Radio Heads (RRHs), our solution completely eliminates the need for dedicated Roadside Units (RSUs) and additional onboard sensors, significantly reducing both system complexity and infrastructure costs, as well as substantially lowering the carbon footprint~\cite{jones,carbs}.

A core contribution of \emph{\papername{}} is its novel adaptive communication-computation synergy, enabled by a lightweight PID-based controller that dynamically modulates the execution frequency of the computationally intensive MUSIC algorithm in response to real-time variations in Signal-to-Noise Ratio (SNR) and vehicle velocity. This dynamic adaptation ensures high and reliable localization accuracy, consistently maintaining worst-case positioning errors below 0.3~m, while achieving notable average energy savings of approximately 43.09\% compared to non-adaptive systems employing vanilla MUSIC, ESPRIT, and Root-MUSIC.
Comprehensive experimental evaluations conducted on an NVIDIA Jetson AGX Xavier platform across multiple operational modes confirm the robustness and effectiveness of our adaptive strategy. Moreover, design space exploration further highlights optimal operating points that balance energy efficiency with stringent accuracy requirements, underscoring the scalability and practical viability of our solution. In addition to energy and accuracy gains, \emph{ACCESS-AV} achieves a substantial $130\times$ reduction in onboard sensor costs aiding to overall sustainability.


Looking ahead, several promising avenues remain open for exploration. These include investigating non-line-of-sight conditions~\cite{nlos}, integrating alternative localization methods such as Received Signal Strength (RSS)~\cite{rss}, Time of Arrival (ToA)~\cite{toa}, and Time Difference of Arrival (TDoA)~\cite{tdoa}, and exploring other AoA techniques, like DeepMUSIC~\cite{deepmusic} and Deep Root-MUSIC~\cite{deeproot}. Further research into more advanced controller designs~\cite{mpc,nn}, extensive hardware platform evaluations~\cite{perceptin,asic}, and cross-layer optimization frameworks~\cite{crx} would also enhance system performance. Crucially, future work will focus on developing comprehensive strategies to enhance reliability, explicitly addressing potential sensor or communication failures~\cite{sense}, and augmenting our solution with additional safety mechanisms. Additionally, quantifying the precise carbon footprint reductions~\cite{cc,cnw} will be vital steps toward achieving sustainable, reliable, and scalable autonomous localization solutions.
\section{Acknowledgements}
\label{sec:ack}
We would like to thank Alish Kanani, Tamoghno Das, Dongjoo Seo, Hyunwoo Oh, Yuxin (Myles) Liu, and Sangeetha Abdu Jyothi for their valuable input at various stages during the preparation of this manuscript. This work was partially supported by the Bob and Barbara Kleist Endowed Graduate Fellowship 2024--25. 
\settopmatter{printacmref=false} 
\bibliographystyle{unsrt}

\vspace{10cm}
\appendix

\section*{Appendix: Acronyms}
\label{appendix:acronyms}
This table provides definitions for the acronyms used throughout this paper.

\setlength{\tabcolsep}{2pt}      
\renewcommand{\arraystretch}{0.9} 

\begin{table}[h!]
\centering
\scriptsize  
\begin{adjustbox}{max width=\columnwidth}
\begin{tabular}{p{1.2cm} p{2.2cm} | p{1.2cm} p{2.2cm} | p{1.2cm} p{2.2cm} | p{1.2cm} p{2.2cm}}
\toprule
\textbf{Acronym} & \textbf{Full Form} & \textbf{Acronym} & \textbf{Full Form} & \textbf{Acronym} & \textbf{Full Form} & \textbf{Acronym} & \textbf{Full Form} \\
\midrule
3GPP    & 3rd Generation Partnership Project   & FFT     & Fast Fourier Transform                      & PBCH    & Physical Broadcast Channel              & SVD     & Singular Value Decomposition \\
5G NR   & 5G New Radio                           & GNSS    & Global Navigation Satellite System          & PID     & Proportional-Integral-Derivative        & UE      & User Equipment \\
ADV     & Autonomous Delivery Vehicle            & GPS     & Global Positioning System                   & PCI     & Physical Cell Identity                  & ULA     & Uniform Linear Array \\
AMC     & Adaptive Modulation and Coding         & IFFT    & Inverse Fast Fourier Transform              & PRB     & Physical Resource Block                 & V2I     & Vehicle-to-Infrastructure \\
AoA     & Angle of Arrival                       & IMU     & Inertial Measurement Unit                   & PSS     & Primary Synchronization Signal          & V2X     & Vehicle-to-Everything \\
AoH     & Angle of Heading                       & LiDAR   & Light Detection and Ranging                 & RF      & Radio Frequency                         & WT      & Wake-Up Time \\
AoIS    & Angle of Incoming Signal               & LoS     & Line of Sight                             & RRH     & Remote Radio Head                       & gNB     & gNodeB \\
AWGN    & Additive White Gaussian Noise          & MIB     & Master Information Block                  & RSU     & Roadside Unit                          & t$_\text{base}$ & Base Wake-Up Time (Hardware Constrained) \\
CP      & Cyclic Prefix                          & MPSoC   & Multiprocessor System-On-Chip             & SNR     & Signal-to-Noise Ratio                     & t$_\text{max}$  & Maximum Allowable Wake-Up Time \\
DoA     & Direction of Arrival                   & MUSIC   & Multiple Signal Classification             & SSB     & Synchronization Signal Block            &         &  \\
EVD     & Eigenvalue Decomposition               & OFDM    & Orthogonal Frequency Division Multiplexing  & SSS     & Secondary Synchronization Signal       & t$_\text{theoretical}$  & Theoretical Baseline \\
\bottomrule
\end{tabular}
\end{adjustbox}
\end{table}

\end{document}